\documentclass[10pt,conference]{IEEEtran}

\usepackage{amsmath,amssymb,amsfonts}
\usepackage{algorithmic}
\usepackage{graphicx}
\graphicspath{ {./graphics/} }
\usepackage{textcomp}
\usepackage{xcolor}
\usepackage{tabularx}
\usepackage[hidelinks]{hyperref}
\usepackage{amsmath}

\usepackage{makecell}
\usepackage{cleveref}
\usepackage{xcolor}
\usepackage{xspace}

\usepackage[most]{tcolorbox}
\usepackage{fancybox}
\usepackage{ifthen}
\newboolean{showcomments}
\setboolean{showcomments}{true}
\ifthenelse{\boolean{showcomments}}
{\newcommand{\nb}[2]{
  \fcolorbox{black}{yellow}{\bfseries\sffamily\scriptsize#1}
{\sf\small$\blacktriangleright$\textit{#2}$\blacktriangleleft$}
 }
 
}
{\newcommand{\nb}[2]{}
 
}
\usepackage{tcolorbox}
\newtcolorbox{mybox}{colback=red!5!white,colframe=red!75!black}
\usepackage{pgfplots}
\pgfplotsset{compat=1.18}

\usepackage{booktabs}

\definecolor{myBlue}{RGB}{0,114,178}
\colorlet{blue}{myBlue}

\newcommand\phase[1]{\tikz[baseline=(X.base)]\node [draw=myBlue,fill=myBlue,thick,rectangle,inner sep=2pt, rounded corners=2pt](X){\color{white}\textbf{#1}};}

\usepackage[framemethod=TikZ]{mdframed}
\mdfsetup{nobreak=false}
\newenvironment{Problem}[1][]{%
  \ifstrempty{#1}%
  {\mdfsetup{%
    frametitle={%
      \tikz[baseline=(current bounding box.east),outer sep=0pt]
      \node[line width=0pt,anchor=east,rectangle,draw=white,fill=white]
    ;}}
  }%
  {\mdfsetup{%
    frametitle={%
      \tikz[baseline=(current bounding box.east),outer sep=0pt]
      \node[anchor=east,rectangle,draw=white,fill=white]
    {\strut #1};}}%
  }%
  \mdfsetup{innertopmargin=-5pt,linecolor=black,%
            linewidth=0.5pt,topline=true,%
            frametitleaboveskip=\dimexpr-\ht\strutbox\relax,skipabove=\topskip,skipbelow=\topskip}
  \begin{mdframed}[nobreak=false]\relax
  }{\end{mdframed}}

\makeatletter
\newcommand{\linebreakand}{%
  \end{@IEEEauthorhalign}
  \hfill\mbox{}\par
  \mbox{}\hfill\begin{@IEEEauthorhalign}
}
\makeatother

\usepackage{subcaption}

\usepackage[framemethod=TikZ]{mdframed}
\mdfsetup{nobreak=false}
\newenvironment{Answer}[1][]{%
  \ifstrempty{#1}%
  {\mdfsetup{%
    frametitle={%
      \tikz[baseline=(current bounding box.east),outer sep=0pt]
      \node[line width=0pt,anchor=east,rectangle,draw=white,fill=white]
    ;}}
  }%
  {\mdfsetup{%
    frametitle={%
      \tikz[baseline=(current bounding box.east),outer sep=0pt]
      \node[anchor=east,rectangle,draw=white,fill=white]
    {\strut #1};}}%
  }%
  \mdfsetup{innertopmargin=-5pt,linecolor=black,%
            linewidth=0.5pt,topline=true,%
            frametitleaboveskip=\dimexpr-\ht\strutbox\relax,skipabove=\topskip,skipbelow=\topskip}
  \begin{mdframed}[nobreak=false]\relax
  }{\end{mdframed}}

\usepackage[switch]{lineno}
\begin{document}
\thispagestyle{plain}
\pagestyle{plain}


\title{Domain-Knowledge Driven Configuration\\Tuning of Kubernetes Workloads}

\author{

\IEEEauthorblockN{Federico~Imberti\textsuperscript{\textsection}}
\IEEEauthorblockA{Akamas\\
Milano, Italy\\
federico.imberti@akamas.io}
\thanks{Manuscript received...}
\and
\IEEEauthorblockN{Luca~Cavagnis\textsuperscript{\textsection}}
\IEEEauthorblockA{Akamas\\
Milano, Italy\\
luca.cavagnis@akamas.io}
\and
\IEEEauthorblockN{Stefano~Cereda}
\IEEEauthorblockA{Akamas\\
Milano, Italy\\
stefano.cereda@akamas.io}
\linebreakand
\IEEEauthorblockN{Giovanni~Gibilisco}
\IEEEauthorblockA{Akamas\\
Milano, Italy\\
giovanni.gibilisco@akamas.io}
\and
\IEEEauthorblockN{Claudio~Menghi}
\IEEEauthorblockA{University of Bergamo\\
Bergamo, Italy\\
McMaster University\\
 Hamilton, Canada}
\IEEEauthorblockA{claudio.menghi@unibg.it}
}

\newcommand{\AKAMAS}{\textsc{Akamas}\xspace}

\newcommand{\numberOfConfigurationRules}{65\xspace}
\newcommand{\numberOfWkldRules}{36\xspace}
\newcommand{\numberOfClusterRules}{9\xspace}
\newcommand{\numberOfNodeRules}{5\xspace}
\newcommand{\numberOfApplicationRules}{15\xspace}
\newcommand{\numberOfReliabilityRules}{27\xspace}
\newcommand{\numberOfComplianceRules}{11\xspace}
\newcommand{\numberOfEfficiencyRules}{15\xspace}
\newcommand{\numberOfStructuralRules}{12\xspace}

\newcommand{\numLinesCodeAlgoHeur}{1\,200\xspace}
\newcommand{\numLinesCodeSMTSol}{500\xspace}

\newcommand{\numKubernetes}{eight\xspace}
\newcommand{\numWorkloads}{14\,742\xspace}

\newcommand{\avgNumViolatedConstraintsBase}{1\,658.10\xspace}
\newcommand{\stdNumViolatedConstraintsBase}{1\,104.25\xspace}
\newcommand{\minNumViolatedConstraintsBase}{157\xspace}
\newcommand{\maxNumViolatedConstraintsBase}{3047\xspace}

\newcommand{\avgNumViolatedConstraintsHeu}{1\,217.24\xspace}
\newcommand{\stdNumViolatedConstraintsHeu}{1\,298.81\xspace}
\newcommand{\minNumViolatedConstraintsHeu}{5\xspace}
\newcommand{\maxNumViolatedConstraintsHeu}{3\,812\xspace}

\newcommand{\avgNumViolatedConstraintsZ}{938.33\xspace}
\newcommand{\stdNumViolatedConstraintsZ}{1\,580.90\xspace}
\newcommand{\minNumViolatedConstraintsZ}{0\xspace}
\newcommand{\maxNumViolatedConstraintsZ}{4\,198\xspace}

\newcommand{\avgCostsBase}{48\,245.63 USD/mo\xspace}
\newcommand{\stdCostsBase}{22\,485.13 USD/mo\xspace}
\newcommand{\minCostsBase}{24\,319.78 USD/mo\xspace}
\newcommand{\maxCostsBase}{90\,262.50 USD/mo\xspace}

\newcommand{\avgCostsHeu}{36\,165.98 USD/mo\xspace}
\newcommand{\stdCostsHeu}{19\,200.30 USD/mo\xspace}
\newcommand{\minCostsHeu}{16\,641.48 USD/mo\xspace}
\newcommand{\maxCostsHeu}{69\,879.24 USD/mo\xspace}

\newcommand{\avgCostsZ}{32\,005.66 USD/mo\xspace}
\newcommand{\stdCostsZ}{16\,026.45 USD/mo\xspace}
\newcommand{\minCostsZ}{15\,830.24 USD/mo\xspace}
\newcommand{\maxCostsZ}{61\,238.45 USD/mo\xspace}

\newcommand{\avgTimeToSolHeu}{0.71 seconds\xspace}
\newcommand{\stdTimeToSolHeu}{0.58 seconds\xspace}
\newcommand{\minTimeToSolHeu}{0.07 seconds\xspace}
\newcommand{\maxTimeToSolHeu}{1.27 seconds\xspace}

\newcommand{\avgTimeToSolZ}{45.54 seconds\xspace}
\newcommand{\stdTimeToSolZ}{34.13 seconds\xspace}
\newcommand{\minTimeToSolZ}{6.64 seconds\xspace}
\newcommand{\maxTimeToSolZ}{75.74 seconds\xspace}

\newcommand{\savingSMTAbsolute}{\textsc{{15\,874} USD}/mo\xspace}
\newcommand{\savingSMTPercentage}{\textsc{32.90\%}\xspace}
\newcommand{\savingHeuristicAbsolute}{\textsc{{11\,591} USD}/mo\xspace}
\newcommand{\savingHeuristicPercentage}{\textsc{24.03\%}\xspace}
\newcommand{\savingDiffHeuZAbsolute}{\textsc{4283} USD/mo\xspace}
\newcommand{\savingDiffHeuZPercentage}{\textsc{31.19\%}\xspace}
\newcommand{\diffTimeToSolHeuZAbsolute}{44.83 seconds\xspace}
\newcommand{\diffTimeToSolHeuZPercentage}{193.86\%\xspace}

\newcommand{\diffConstrBrokenHeuBaseAbsolute}{440.86\xspace}
\newcommand{\diffConstrBrokenHeuBasePercentage}{30.68\%\xspace}
\newcommand{\diffConstrBrokenZBaseAbsolute}{719.77\xspace}
\newcommand{\diffConstrBrokenZBasePercentage}{55.47\%\xspace}
\newcommand{\diffConstrBrokenHeuZAbsolute}{278.91\xspace}
\newcommand{\diffConstrBrokenHeuZPercentage}{25.89\%\xspace}

\maketitle
\begingroup\renewcommand\thefootnote{\textsection}
\footnotetext{This work has been done while Federico Imberti and Luca Cavagnis were master's students at the University of Bergamo, Italy.}

\begin{abstract}
Industrial software applications often require engineers to tune the software deployment platform.
Configuration tuning identifies valuable configurations that minimize costs while ensuring the application performance.
However, finding these configurations is difficult. 
The number of tunable configuration parameters in industrial applications is large, and engineers cannot assess different configurations in production because users may experience performance degradation.
This work targets Kubernetes: A system for automating the deployment, scaling, and management of containerized applications.

This paper proposes a domain-knowledge-driven solution for finding configurations of industrial applications in a practical time.
The solution relies on a catalog of constraints encountered by domain experts in similar applications. 
Depending on the problem under analysis, engineers can customize the optimization problem by selecting the relevant constraints from the catalog for the application at hand.
We rely on two alternative solvers to find an optimal configuration: The first is based on heuristics, while the second reuses existing optimization-based solutions.
We implemented our solution within an industrial tool from a software company specializing in performance engineering and optimization. 
We compare our solution to a baseline, considering a set of predefined configuration values provided by engineers.
Our results confirm the effectiveness of our solution: Compared to the baseline, the heuristic-based and optimization-based solutions provide significant average savings of \savingHeuristicAbsolute and \savingSMTAbsolute per Kubernetes, respectively, accounting for \savingHeuristicPercentage and \savingSMTPercentage of the costs, in practical time (\avgTimeToSolHeu per extraction for the heuristic-based and \avgTimeToSolZ per extraction for the optimization-based solution).
\end{abstract}


\section{Introduction}
Deployment of industrial software is complicated. 
It typically requires engineers to configure their applications and platforms: A complex, time-consuming activity widely recognized by the software engineering community~\cite{chen2018femosaa,chen2024adapting,han2016empirical,nair2018finding}.
Modern software applications typically offer a large set of configuration options. 
The running platform (typically from third parties) is also configurable, enabling engineers to select various resource types (e.g., computational and storage) depending on the application's needs.
Configuration options have a significant impact on software
performance~\cite{chen2018femosaa,chen2024adapting,han2016empirical,nair2018finding}. 
Recent surveys~\cite{sayagh2018software,10.1145/3702986} provide sets of challenges related to these activities.

Performance optimization aims to identify a configuration that minimizes costs while ensuring availability and performance requirements, often specified by Key Performance Indicators (KPIs)~\cite{ardagna2014multi} or Service Level Agreements (SLA) determined with the customer (for instance, an upper bound in the latency of the service).
Performance optimization is a widely studied problem by the academic  (e.g.,~\cite{gong2024dividable,cheng2023hinnperf,gong2023predicting,ha2019deepperf,temple2019towards}) and industrial (e.g.,~\cite{stefano2023method,cereda2022method}) software engineering communities.
For example, our industrial partner (\AKAMAS~\cite{Akamas}) is specialized in the optimization of applications, including the infrastructure where they are deployed.

Traditional manual configuration tuning methods struggle to cope with the dynamic nature of modern infrastructures that have a large set of configuration options~\cite{didona2015enhancing}. 
For example,
Hadoop has about 900 parameters
across four sub-systems~\cite{krishna2020conex}.
Therefore, automated performance tuning approaches have gained significant attention~\cite{gong2024dividable,cheng2023hinnperf,gong2023predicting,ha2019deepperf,temple2019towards}. 
For example, Bayesian Optimization~\cite{agarwal2019active,hsu2018arrow} has been recently proposed for efficiently exploring large configuration spaces in the absence of an explicit optimization function.
These techniques have also been applied in industrial contexts. 
For example, \AKAMAS~\cite{Akamas_product} relies on Bayesian Optimization and contextual forecasting in some of its offerings to adapt the configuration after environmental changes and improve its robustness~\cite{Cereda:2022aa}. This technology led to significant practical cost reductions: It reduced the costs of a gambling company (Sisal~\cite{sisalAbout}) and a travel technology company (Sabre~\cite{sabreAbout}) by 58\%~\cite{akamasSisal} and 50\%~\cite{akamasSabre}.

Automated configuration tuning is \emph{computationally intensive}. A recent software engineering review~\cite{10.1145/3702986} observed that random sampling is the most commonly used method, and manual sampling is still largely used.
Recent studies reported that measuring a single configuration on the database system MARIADB takes more than two hours~\cite{chen2021multi,chen2024adapting}.
Many solutions have tried to address this problem, mostly by considering surrogate models (e.g.,~\cite{BOCA, atconf, flash, Ottertune, restune, ROBOTune, SMAC, bestconfig, lopez2016irace, gga, paramils, muhlbauer2023analyzing, Sway, conex}).
This problem is exacerbated in early configuration phases, when performance engineers need to identify a configuration in (a)~a limited time (less than one hour), and (b)~without any historical data on the application's behavior. 
For example, a recent incident involved a global leader in digital services and real-time transactional platforms in the retail market~\cite{akamasInsightPrevention}. This company manages a complex technology landscape, supporting millions of customer interactions daily across thousands of physical and digital endpoints.
They recently experienced a production outage that severely impacted one of their most business-critical applications.
The incident occurred shortly after a new software release. While the release itself did not contain breaking changes, it exposed a pre-existing fragility in the cluster’s configuration: Memory usage had been silently increasing over time across multiple nodes. 
Our industrial partner (\AKAMAS) has a solution based on Bayesian Optimization relying on Gaussian Processes (GP) as probabilistic surrogate models to search for optimal configurations in their  offering~\cite{akamasOfflineAbout}, which handles these types of problems.
However, despite its effectiveness, this solution does not apply to early configuration phases, as it takes hours (or days) to find a (sub-)optimal configuration based on historical performance data. 
Moreover, it is a ``black box'' solution, often lacking explainability, which limits its adoption in critical production environments where human operators must understand and trust optimization decisions.

To address these problems, this work proposes \emph{domain-driven configuration tuning}.
The general idea of our approach is to reuse the domain knowledge that engineers have accumulated over time to configure new systems.
Specifically, our domain-driven configuration tuning solution relies on two pillars: A \emph{catalog of configuration rules} and a \emph{solver} that resolves the configuration problem. 
The catalog of configuration rules contains configuration patterns expressed as inequalities.
These rules were generated by the years of experience accumulated by \AKAMAS on the tuning of systems for their customers.
They describe (and formalize) recurrent configuration problems.
When performance engineers analyze a new system, they can import the configuration rules relevant to the problem at hand from the catalog.
This activity generates a configuration problem that must be solved using an appropriate solver.
In this work, we propose two solvers for the configuration problem: One is heuristic-based, and the other is optimization-based.
The heuristic-based solver identifies a (sub-)optimal configuration for a system using hard-coded algorithms.
However, since this solution is hard-coded, the performance engineer must manually edit the software's code every time new rules are added (or removed). 
Unlike it, the optimization-based solver uses optimization algorithms (e.g., simplex~\cite{z3ArithOptimAbout}) to find a (sub-optimal) configuration. 
Depending on the configuration rules selected by the performance engineer, the optimization solver compiles a new optimization problem, which is then solved using an existing solver (e.g., the Z3 SMT solver~\cite{demoura2008z}).
Therefore, the optimization-based solution enables performance engineers to add or remove rules that are automatically injected into or extracted from the optimization problem without requiring any modification of the algorithms used to solve the problem, as these tasks can be done easily from the user interface.
Despite the proposed approach and solution being general and applicable to many performance engineering problems, we propose an implementation targeting Kubernetes (see~\ref{sec:Kubernetes}), i.e., a system for automating deployment, scaling, and management of containerized applications.

To evaluate our solution, we considered two industrial applications from two large industrial players supported by our partner \AKAMAS, in the configuration of their software platform.
They are two large applications used by millions of customers each year.
In total, we considered \numWorkloads workloads deployed over \numKubernetes clusters of nodes. 
We used our domain-driven configuration tuning solution to find valuable configurations and compare them with a baseline consisting of values extracted from the targeted Kubernetes before any intervention.
Our results confirm the effectiveness of our solution: Compared to the baseline, the heuristic-based and optimization-based solutions provide significant average savings of \savingHeuristicAbsolute and \savingSMTAbsolute per Kubernetes, respectively, accounting for \savingHeuristicPercentage and \savingSMTPercentage of the costs, in practical time (with an average of \avgTimeToSolHeu per extraction for the heuristic and \avgTimeToSolZ per extraction for the optimization-based solution).
The solutions presented in this work have been integrated into the \textsc{Insights} commercial tool~\cite{Insights}, now commercialized by \AKAMAS.

This paper is organized as follows. 
\Cref{sec:background} presents background notions.
\Cref{sec:Problem} defines the problem considered in this work.
\Cref{sec:solution} illustrates our solution.
\Cref{sec:Evaluation} evaluates the proposed solution.
\Cref{sec:RelatedWork} discussed related works.
\Cref{sec:Conclusions} concludes this work. 
\section{Background}
\label{sec:background}
This section provides background relevant knowledge on Kubernetes (\Cref{sec:Kubernetes}), the system targeted by our implementation, and Satisfiability Modulo Theory (SMT) and Z3 (\Cref{sec:sat}), the technology used by the optimization-based solution to find a valuable configuration.

\subsection{Kubernetes}
\label{sec:Kubernetes}
Kubernetes~\cite{kubeAbout} was initially developed by Google~\cite{googleKube} and is currently managed by the Cloud Native Computing Foundation (CNCF)~\cite{cncfAbout}. 
It is an open-source extensible platform for deploying, scaling, and managing containerized applications. 
Kubernetes deploys complex applications (consisting of many containers) by defining YAML and JSON files describing their structure.
A Kubernetes architecture is a cluster of nodes. An application (or distributed service) runs on this cluster. 
We define the application by using one or more workloads. 
A workload defines how the application should run. 
For this reason, we decided to operate on workloads instead of the nodes or cluster configuration. 
Furthermore, operating at a lower level of abstraction enables more granular control over the tuning, allowing us to configure the number of millicores of each CPU and the number of MiB (mebibytes) of each memory unit of each workload.
Instead of controlling and deploying each container separately, the engineer provides YAML and JSON specification files. 
Kubernetes uses this specification to run the application continuously, reconciling the cluster's status to the desired specification. 
For example, a configuration file could specify Kubernetes to run three copies of a Docker image and expose it on port 80.
The Kubernetes controller will deploy three containers from the given image and expose them on port 80. 
If a container crashes, the controller deploys a new one and frees the resources of the failed container. 
In short, Kubernetes enables engineers to control the containers of a distributed application, ensuring an appropriate amount of resources is present, via API calls and specification files.

\subsection{SMT and Arithmetical Optimization with Z3}
\label{sec:sat}
Satisfiability modulo theories (SMT)~\cite{barrett2018satisfiability} concerns the problem of verifying whether a formula is satisfiable. 
The formula can be mathematical and involve many different elements, such as real numbers, integers, arrays, and bit vectors. 
These elements can be considered within the formula satisfiability problem (``modulo'') a formal theory (typically the first-order logic with equality).
The Z3 SMT solver~\cite{Z32020,de2008z3} is a tool for SMT satisfiability. 
It is compatible with the STMLIB format~\cite{smllibfAbout}, which describes the SMT-LIB input/output language for SMT solvers and its semantics.
It has recently been extended to solve optimization problems, relying on the primal Simplex~\cite{z3ArithOptimAbout} algorithm.
This algorithm is a standard solution for solving optimization problems, which is largely used in the research literature (e.g.,~\cite{Orlin1997, HELGASON199585, doi:10.1287/moor.4.1.39}).
In this work, we consider Z3 because it is an award-winning~\cite{ETAPS2018,SIGPLAN2020} industry-strength tool supported by Microsoft~\cite{z3MicrosftSupportAbout}.

\section{Problem Definition}
\label{sec:Problem}

Customers typically contact performance engineers to configure their applications and platforms.
At the early configuration stages, the performance engineers need to face two challenges: (a)~they have to identify a valuable configuration in a short amount of time, to provide the customer with prompt feedback; and (b)~they do not possess any historical data on the application’s behavior other than its current state.

Two approaches are currently used as standard practices to address these challenges.

\emph{Manually selecting the configuration}. 
Performance engineers can use their domain knowledge and experience to select a set of values for the configurable parameters that are adequate for the specific domain and application~\cite{pepperdataManualConfTun}.
This approach has significant limitations since it is open to many fallacies~\cite{akamasLoweringGcDangerous}~\cite{pepperdataManualConfTunFails}.
First, engineers must consider many constraints when selecting values for their configuration parameters. 
For example, performance engineers might have requirements related to the lower bound value for the memory allocated to the application that, if not updated to match new requirements, could lead to outages~\cite{akamasInsightPrevention}.
To satisfy these requirements, engineers might be tempted to overestimate the memory, which in turn may violate other related requirements, for instance by exceeding the memory upper bound.
Consequently, engineers are likely to involuntarily select values violating one or more constraints of the application while trying to optimize a part of it.

\emph{Automatically selecting the configuration}.
Performance engineers might use automated tools to select the configuration values. 
For example, one of the tools provided by our industrial partner (\AKAMAS~\cite{Akamas}) uses Bayesian optimization to find an optimal configuration by considering historical application data and domain-specific constraints on the application~\cite{Akamas_product}.
However, this solution is not adequate for our usage scenario as it requires historical application data and can take up to days to converge to an optimal configuration.

Therefore, the goal of this work is to address the following problem:
\begin{Problem}[Problem]
Find an adequate configuration in a practical time for a given Kubernetes application. 
\end{Problem}
With ``adequate'' we mean a configuration that minimizes the cost of the cluster while violating the fewest number of constraints on the configuration values.

To address this problem, we propose \emph{Domain-Knowledge-Driven Configuration Tuning}.

\section{Domain-Knowledge-Driven Configuration Tuning}
\label{sec:solution}

This section provides an overview of the proposed solution (\Cref{sec:overview}), presents the catalog of configuration rules (\Cref{sec:catalog}), and the heuristic-based (\Cref{sec:INSIGHTS}) and optimization-based (\Cref{sec:ZINSIGHTS}) solvers.
Finally, it outlines our tool support
(\Cref{sec:userExperience}).

\begin{figure}[t]
    \centering
    \tikzstyle{output} = [coordinate]

\begin{tikzpicture}[auto,
 block/.style ={rectangle, draw=black, thick, fill=white!20, text width=6em,align=center, rounded corners},
 block1/.style ={rectangle, draw=blue, thick, fill=blue!20, text width=7em,align=center, rounded corners, minimum height=5em},
 line/.style ={draw, thick, -latex',shorten >=2pt},
 cloud/.style ={draw=red, thick, ellipse,fill=red!20,
 minimum height=1em}]

\draw (0,0) node[block,node distance=2.1cm,minimum height=5.5em] (B1) {};

\node[block, right of=B1,node distance=4.5cm,minimum height=5.5em] (B2){};

\node[block, below of=B1,text width=5em,node distance=0cm,draw=white!80,] (B6){\phase{1} \footnotesize Catalog of Configuration Rules};

\node [output, right of=B2,node distance=3cm] (ENDNode) {};

\node[block, below of=B2,text width=4em,node distance=-0.45cm,draw=white!80,] (B5){\phase{2}\phantom{x}\footnotesize Solver};
\node[block, below of=B2,text width=5em,node distance=0cm] (B3){\footnotesize Heuristic};

\node[block, below of=B3,text width=5em,node distance=0.545cm] (B4){\footnotesize Optimization};

\draw[-stealth, text width=6em,align=center] (B1.east) -- (B2.west)
    node[midway,above]{\footnotesize Configuration Problem};
\draw[-stealth, text width=6em,align=center] (B2.east) -- (ENDNode.west)
    node[midway,above]{\footnotesize Configuration Values};


 \end{tikzpicture}
    \caption{Domain-Knowledge Driven Configuration Tuning.}
    \label{fig:dkdct}
\end{figure}
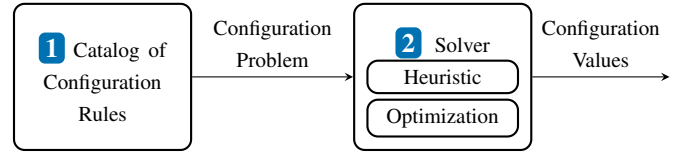

\subsection{Overview of the Approach}
\label{sec:overview}

\Cref{fig:dkdct} presents our domain-knowledge-driven configuration tuning solution.
Our solution relies on two components: The \emph{Catalog of Configuration Rules} (\phase{1}) and the \emph{Solver} (\phase{2}).

\begin{table*}[t]
\caption{Configuration Rules from the Catalog.}
    \label{tab:rules}
\begin{tabular}{l p{4.5cm} l}
 \toprule
   \textbf{ID}  &     \textbf{Explanation} &    \textbf{Expression} \\
\midrule
\#1 & The ratio of CPU limits to CPU requests is too high, which can lead to application and cluster stability risks & \makecell[l]{$@lr\_ratio\_cpu > MAX\_CPU\_LIM\_REQ\_RATIO$} \\ 
\midrule
\#2 & Assign the CPU limits based on the throttling value & \makecell[l]{$limits\_cpu\_p99  <  (@throttling\_cpu==1)*(limits\_cpu\_p99 + cpu\_throttled\_p99) $\\
\phantom{$limits\_cpu\_p99  <XX$}$+(@throttling\_cpu==0)*( \$requests\_cpu\_p99 * @old\_lr\_ratio\_cpu) $  
}\\
\midrule
\#3 & Memory limit insufficient given heap and off-heap requirements & \makecell[l]{$ \$limits\_memory\_p99 <  (\$heap\_size + \$off\_heap\_size)/PROBLEMS\_SLACK $} \\
\bottomrule
\end{tabular}
\end{table*}

\emph{Catalog of Configuration Rules} (\phase{1}). 
The catalog of configuration rules collects configuration rules accumulated by domain experts from \AKAMAS, leveraging years of experience in configuration tuning.
These rules represent recurrent problems that performance engineers addressed when configuring software applications and their platforms.

This approach enables engineers to reuse domain knowledge and prevent constraint overlooking (e.g., the memory upper bound).
Solving the configuration problem requires finding a set of configuration values that satisfies all constraints (whenever possible) or minimizes the violated constraints.
Both our heuristic and optimization-based approaches use the catalog.
However, only the optimization-based solution builds an optimization problem using them as constraints and allows the user to activate and deactivate the constraints from the user interface.

 \emph{Solver} (\phase{2}). The solver searches for a set of configuration values that satisfies all constraints (whenever possible) or minimizes the violated constraints.
This work proposes a heuristic-based and an optimization-based solver.
They are complementary: the first relies on a set of hard-coded heuristics that are likely to be effective, but are hardly generalizable to different configuration problems (They require performance engineers to modify the source code itself), and the second is more generic and likely to be less effective. 
These two solutions will be compared in our evaluation (\Cref{sec:Evaluation}).

\subsection{Catalog of Configuration Rules}
\label{sec:catalog}

The catalogue contains \numberOfConfigurationRules configuration rules.
Industrial experts have collected them over the years they spent solving hundreds of configuration problems. The rules act at three different levels: Workload (\numberOfWkldRules), cluster (\numberOfClusterRules), node (\numberOfNodeRules), and application (\numberOfApplicationRules).
In this work, we consider the workload level because it supports a finer granularity for our tuning efforts. 
It allows us to focus on each CPU and each memory unit independently from the others in the node and cluster. 
The approach can be generalized to operate at the cluster, node, or application level.
The rules refer to reliability (\numberOfReliabilityRules), compliance (\numberOfComplianceRules), and efficiency (\numberOfEfficiencyRules).
Reliability opportunities allow setting values that will not crash the system, for instance, defining a minimum number of MiB that can be assigned to the memory. Compliance opportunities define best practices in configuring Kubernetes environments.
Efficiency opportunities identify ways to reduce resource usage and costs. 
Additionally, there are \numberOfStructuralRules rules classified as ``structural'' which we only need for the optimization-based solution to identify a mathematically feasible solution that is also within the domain of each tunable variable. 
Since these rules are required to solve the optimization problem, they cannot be deactivated. 

\Cref{tab:rules} reports three rules used for illustration purposes. 
The complete list of \numberOfConfigurationRules rules is not shared with this publication since it is proprietary.
The rules are tagged with an identifier (ID), they contain an explanation (Explanation), and they have an arithmetic relation (Expression).
The identifier is used to recognize the different rules.
The explanation describes their usage scenario and goal.
For example, the rule \#1 specifies that ``\emph{The ratio of CPU limits to CPU requests is too high, which can lead to application and cluster stability risks}''.
The expression contains the arithmetic relation that defines constraints configuration values must respect.
Constants are identified in upper case.
Variables preceded by the symbol ``\$'' represent configurable values.
Variables preceded by the symbol ``@'' represent aliases (placeholders made up of variables and constants used to write more concise expressions) with a definition stored in a dedicated file.
Names preceded by the symbol ``\_'' represent a function that will be computed during the parsing; its definition is stored in another file.
With this grammar we can also express conditions, as seen in rule \#2 which requires the value of $ \$limits\_cpu\_p99 $ to be less than $limits\_cpu\_p99 + cpu\_throttled\_p99$ if the condition defined in the alias $@throttling\_cpu$ is true (i.e., $@throttling\_cpu==1$), otherwise (i.e., $@throttling\_cpu==0$) it must be less than $\$requests\_cpu\_p99 * @old\_lr\_ratio\_cpu$.

\subsection{Heuristic-based}
\label{sec:INSIGHTS}
The heuristic-based solution is domain-specific.
Given a configuration problem, engineers programmatically define how to generate configuration values.
These programs define procedures to compute new configuration values from the previous cluster configuration values before any optimization intervention is executed. The solution is articulated in three steps.
\begin{itemize}
    \item[(a)] All the rules from the catalog (\Cref{sec:catalog}) are used to screen the input dataset (i.e., the initial configuration of the system made by the customer) to identify which problems are present. 
    \item[(b)] New configuration values are computed using the baseline values as inputs to a series of proprietary algorithms, which will not be disclosed in this work as they are intellectual property of our industrial partner. 
    These algorithms do not change across applications and can only be modified by editing the program's source code. 
    They are implemented as (approximately) \numLinesCodeAlgoHeur lines of code. 
    Depending on the specific IT stack and configuration, the tuning can be more or less aggressive. 
    Moreover, the tuning must account for the workloads that implements a Java~\cite{javaAbout} component, as it will need to model the memory differently compared to a non-Java workload. 
    In this case, we have to account separately for the heap usage of the Java Virtual Machine (JVM).
    \item[(c)] When the new values have been computed and written in the dataset, the rules in the catalog are used again to screen it to identify which problems are now present.
\end{itemize}
Despite the program solving violations of different rules (and improving the solution if no violations are detected), the rules are not application-specific, as they are not bound to a specific problem.
This solution offers many advantages, since it is fast (i.e., rules explicitly define how to compute new values) and based on hard-coded domain knowledge. Therefore, it is likely to produce adequate configuration values. 

\subsection{Optimization-Based}
\label{sec:ZINSIGHTS}
The optimization-based solution relies on three steps to solve the configuration problem: (a) extracting the configuration problem related to each workload, (b) translating the configuration problem into an optimization problem, and (c) solving the optimization problem.

\begin{itemize}
    \item[(a)] \emph{Extracting the Configuration Problems Related to Each Workload.} 
    For every node and cluster, our solution creates an optimization problem related to each workload of the Kubernetes, considering the rules selected from the catalog (see~\Cref{sec:catalog}).
        
    \item[(b)] \emph{Translating the Configuration Problem into an Optimization Problem}. This step extracts an optimization problem from the configuration problem. The optimization problem has the form:
    \begin{align}
       & \underset{c}{minimize} && f(c) && \label{costFunction}\\  
       & subject\ to &&  g_i(c) \leq 0 && i=1,\ldots m \label{constraint} \\
        & && h_j(c) =0 && j=1,\ldots p \label{constraintTwo}
    \end{align}
    Specifically, the optimization problem aims to detect a configuration ($c$) that minimizes a cost function (Function~\ref{costFunction}), and respects a set of configuration constraints (Relations \ref{constraint} and \ref{constraintTwo}).
    The objective function maximizes expected savings while minimizing resource use (CPU millicores and MiB of memory) and the number of violated constraints. 
    It is a linear objective function. 
    Its specific definition depends on the configuration rules selected from the catalog.
    Each configuration constraint is extracted from a configuration rule.
    Specifically, the arithmetic expression associated with each configuration rule (see \Cref{sec:catalog}) is extracted and added as a constraint ($g_i(c) \leq 0$ or $h_j(c) =0$) to the optimization problem.
    For example, rule \#1 defines a problem (a situation we do not want to happen in our workloads); so the corresponding constraint becomes $@lr\_ratio\_cpu <= MAX\_CPU\_LIM\_REQ\_RATIO$ (a situation we want to happen in our workloads).
    \item[(c)] \emph{Solving the Optimization Problem}. 
    Solving the optimization problem requires selecting an appropriate solver for the optimization problem under analysis. 
    Our solution is general and not bound to a specific solver. Therefore, different solvers can be selected for implementing our solution. Considering that our constraints are linear, we selected the Z3 SMT solver~\cite{Z32020,de2008z3}, since it can solve optimization problems (relying on primal Simplex~\cite{z3ArithOptimAbout}), and it is an award-winning~\cite{ETAPS2018,SIGPLAN2020} industry-strength tool supported by Microsoft~\cite{z3MicrosftSupportAbout}.
    Assessing the implications of this decision is part of our evaluation.
\end{itemize} 
Our optimization-based solution is implemented as \numLinesCodeSMTSol lines of Python code.
We developed a custom grammar, a parser implemented with PyParsing~\cite{pyparsingAbout}, and a series of custom translators (``transformers'') to translate the expressions from text to Polars and Z3 formats.
We use the Python API of Z3 (Z3Py~\cite{z3PyAbout}) to interact with the solver.
We built an add-on to enforce different constraints based on their classification. 
Our add-on enables performance engineers to turn off some problems to reach an admissible solution by using \emph{soft constraints}.
To penalize the deactivation of constraints, we assigned it a relative priority (from 1 to 4).
We used the priority to select how to deactivate constraints, considering that in our scenario, reliability issues are more important to solve than compliance issues, which are more important than efficiency issues.

\subsection{Tool Support}
\label{sec:userExperience}
We implemented our solution within the \textsc{Insights} commercial tool~\cite{Insights}, now commercialized by \AKAMAS. 
\Cref{fig:toolSupport} presents an overview of the interface of \textsc{Insights}.

\AKAMAS enables the performance engineer to first extract readings from the customer's Kubernetes. 
This feature is implemented using third-party observability platforms (e.g., Dynatrace~\cite{dynatraceAbout} and Datadog~\cite{datadogAbout}), which automatically download the information regarding the cluster's configuration. The downloaded data is stored on the machine where the tool is running.

Once the tool has been launched and the user has completed the login, they are prompted to select a reading to use as the baseline among the ones they have downloaded. 
Then, the user is directed to the main dashboard (Figure~\ref{fig:dashboard}).
Before the dashboard is loaded, the heuristic-based and optimization-based solutions are executed in parallel. 
When both solvers terminate, the results are visualized on the dashboard, where either the heuristic-based or the optimization-based data are loaded when the user acts upon the toggle ``Show Optimization-Based Results'' below the header.
The dashboard is divided into three views.
The ``Efficiency'' view summarizes the efficiency gains computed as cost savings, which can be analyzed by each layer (application, workload, etc.), and the variation of allocated CPU cores and memory GiB.
The ``Reliability'' view reports reliability risks for all the layers.
Finally, the ``Compliance'' view reports the issues related to compliance.

Our implementation enables performance engineers to analyze specific workloads. 
\Cref{fig:wkldDetails} shows the interface used to inspect a workload. 
The ``Recommendation'' view displays the new values for the configuration parameters (e.g., CPU and memory requests and limits, the new heap size of the JVM if the workload uses Java). 
The ``Impact'' view shows the expected impact (efficiency, reliability, and compliance) of setting these configuration values on the workload.
An additional perspective (not reported here not to disclose the list of configuration rules) enables the performance engineer to globally (i.e., for all of the optimization problems) activate or deactivate configuration rules using a list of checkboxes. 
\begin{figure*}[th!]
    \begin{subfigure}[b]{\linewidth}
    \centering
\includegraphics[width=\linewidth]{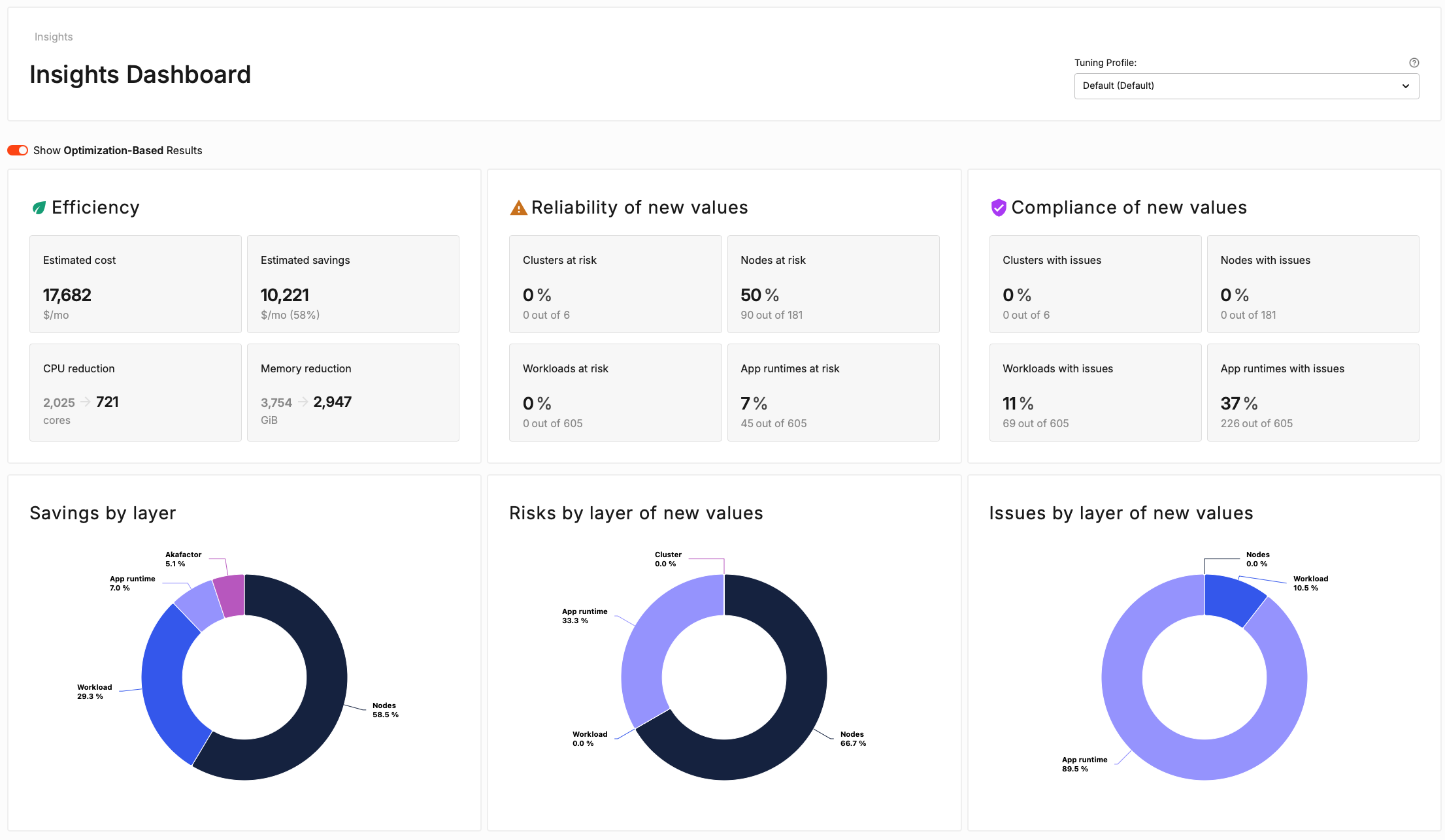}
  \caption{Domain-knowledge-driven configuration tuning: Main dashboard.}
    \label{fig:dashboard}
\end{subfigure}\\
\vspace{0.5cm}
\begin{subfigure}[b]{\linewidth}
    \centering
\includegraphics[width=\linewidth]{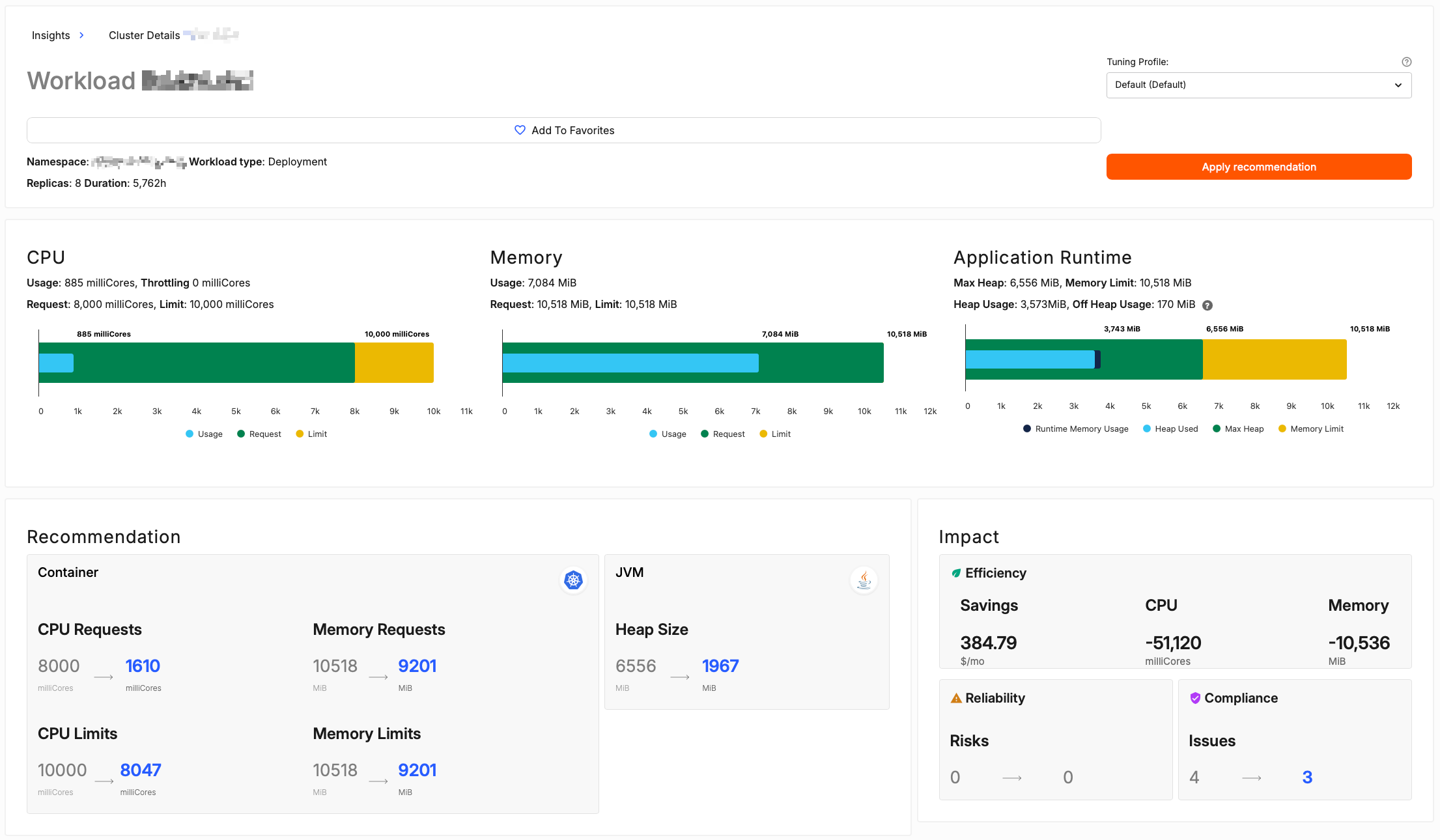}
  \caption{The page detailing a single workload.}
    \label{fig:wkldDetails}
\end{subfigure}
\caption{An overview of the interface of \textsc{Insights}.}
    \label{fig:toolSupport}
\end{figure*}
Some of the problems displayed in this perspective are grayed out and cannot be disabled as they are part of the \numberOfStructuralRules structural constraint that are needed for the optimization solver to find a mathematically feasible solution, as described in~\Cref{sec:overview}. Additionally, the user can toggle on or off rules locally (i.e., for a single workload).


\section{Evaluation}
\label{sec:Evaluation}
Our evaluation considers the following research questions:
\begin{itemize}
    \item \textbf{RQ1 (Effectiveness)}. How \emph{effective} is our solution in finding adequate configurations, and how do the heuristic-based and optimization-based solutions compare?
    (\Cref{sec:Effectiveness})
    \item \textbf{RQ2 (Efficiency)}. How \emph{efficient} is our solution in finding adequate configurations, and how do the heuristic-based and optimization-based solutions compare?
    (\Cref{sec:Efficiency})
\end{itemize}
We present our benchmarks and study subjects (\Cref{sec:Benchmarks}). 
Then, we assess the effectiveness (\Cref{sec:Effectiveness}) and efficiency (\Cref{sec:Efficiency}) of our solution.
Finally, we present threats to validity (\Cref{sec:Threats}) and reflect on our results (\Cref{sec:Discussion}).

\subsection{Benchmarks and Study Subjects}
\label{sec:Benchmarks}
We considered a \emph{benchmark} from two real-world industrial applications.
These applications are used by millions of customers each year.
In total, we considered \numWorkloads workloads deployed over \numKubernetes clusters of nodes.
We considered three \emph{study subjects} for the identification of the configuration values: A baseline solution using a set of values provided by the customer, and our heuristic-based and optimization-based solutions.

\subsection{Effectiveness (RQ1)}
\label{sec:Effectiveness}
We assessed the effectiveness of our solution as follows.

\textbf{Methodology}. We considered our study subjects.
To extract the baseline configuration, we built a custom extractor that downloads the Kubernetes configuration. 
Then, we run our study subjects on each benchmark. 
We recorded the configuration values obtained by each study subject.
We considered two metrics for comparing our study subjects: (a)~the number of violated constraints, and (b)~the costs of the selected configuration.
We considered these metrics for the following reasons.

Violated constraints can represent potential issues that may result in outages that severely impact business-critical applications and significant financial losses (see~\cite{akamasInsightPrevention} for a concrete example).
Therefore, when finding a configuration that satisfies all the constraints is impossible, performance engineers are interested in minimizing the set of violated constraints, prioritizing the critical ones, which are mostly those we classified as ``reliability'' in the Catalog~(see \Cref{sec:catalog}). 
We do not expect all the rules to be violated. 
For instance, it can happen that to satisfy a reliability constraint, we may have to break an efficiency constraint. We are counting on our system to find a good trade-off between reliability and cost. 
Except for the rules we defined as ``structural'', and some other rules having a very high priority, almost all the others are inserted in our optimization problem as \emph{soft constraints} the optimizer can turn off (see~\Cref{sec:ZINSIGHTS}).

Cost is a primary metric for selecting configuration values, since companies do not want to incur unnecessary expenditures.

\textbf{Results}. 
We separately discuss the results related to the number of violated constraints and the configuration costs. 
We considered the statistics related to each optimized Kubernetes.
We estimated the monthly costs~(USD/mo) in United States Dollars~(USD), considering publicly available quotas from lead service providers. 
They provide an average price for the CPU and memory in terms of USD/(millicores*hours) and USD/MiB. 
We used this value to estimate the cost of a single workload. 
This value is a good estimation of the costs.
Indeed, the customer would pay for the nodes in the Kubernetes cluster, which are proportional to the workloads. 

\emph{Number of Violated Constraints}. \Cref{fig:ViolatedConstraints} presents the number of total violated constraints for our baseline, our heuristic-based, and optimization-based solutions. The diamond depicts the average. The average number of violated constraints for our baseline, heuristic-based, and optimization-based solutions is respectively \avgNumViolatedConstraintsBase (\textit{min}=\minNumViolatedConstraintsBase, \textit{max}=\maxNumViolatedConstraintsBase, \textit{StdDev}=\stdNumViolatedConstraintsBase) and \avgNumViolatedConstraintsHeu (\textit{min}=\minNumViolatedConstraintsHeu, \textit{max}=\maxNumViolatedConstraintsHeu, \textit{StdDev}=\stdNumViolatedConstraintsHeu), \avgNumViolatedConstraintsZ (\textit{min}=\minNumViolatedConstraintsZ, \textit{max}=\maxNumViolatedConstraintsZ, \textit{StdDev}=\stdNumViolatedConstraintsZ). 
All the violated constraints are soft: All the returned configurations are feasible despite breaking some constraints as a result of a trade-off between reliability and cost.

The average number of violated configuration rules is significantly reduced by our heuristic-based (-\diffConstrBrokenHeuBaseAbsolute$\approx$-\diffConstrBrokenHeuBasePercentage) and optimization-based solutions (-\diffConstrBrokenZBaseAbsolute$\approx$-\diffConstrBrokenZBasePercentage) compared to the baseline.
There are no cases where the baseline outperforms our heuristic-based and optimization-based solutions. 
The Wilcoxon signed-rank test~\cite{c4091bd3-d888-3152-8886-c284bf66a93a} confirms that there is empirical evidence~($p<0.05$) about the superiority of the heuristic-based and optimization-based solutions compared to the baseline configuration.

The number of violated configuration rules is significantly reduced (-\diffConstrBrokenHeuZAbsolute$\approx$-\diffConstrBrokenHeuZPercentage) by our optimization-based solution compared to the heuristic-based solution.
There are no cases in which the heuristic-based solution outperforms the optimization-based one.
The Wilcoxon signed-rank test~\cite{c4091bd3-d888-3152-8886-c284bf66a93a} confirms that our empirical evidence~($p<0.05$) about the superiority of the optimization-based solution compared to the heuristic-based solution.

\emph{Configuration Costs}. \Cref{fig:Costs} presents the costs for our baseline, heuristic-based, and optimization-based solutions.
The average costs for our baseline, heuristic-based, and optimization-based solutions are respectively \avgCostsBase (\textit{min}=\minCostsBase, \textit{max}=\maxCostsBase, \textit{StdDev}=\stdCostsBase),  
\avgCostsHeu (\textit{min}=\minCostsHeu, \textit{max}=\maxCostsHeu, \textit{StdDev}=\stdCostsHeu), and 
\avgCostsZ (\textit{min}=\minCostsZ, \textit{max}=\maxCostsZ, \textit{StdDev}=\stdCostsZ).

\begin{figure*}[t]
    \begin{subfigure}[b]{0.5\linewidth}
    \centering
    \includegraphics[width=\linewidth]{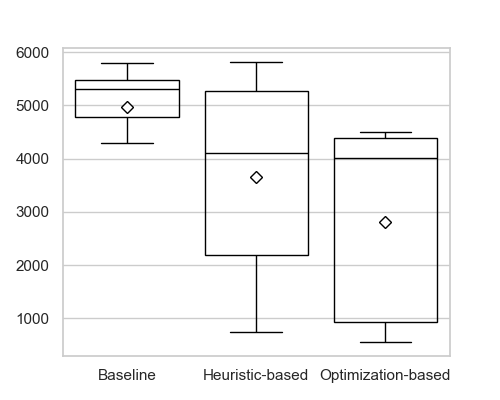}
    \caption{Number of Violated Constraints.}
    \label{fig:ViolatedConstraints}
    \end{subfigure}\hfill
    \begin{subfigure}[b]{0.5\linewidth}
    \centering
    \includegraphics[width=\linewidth]{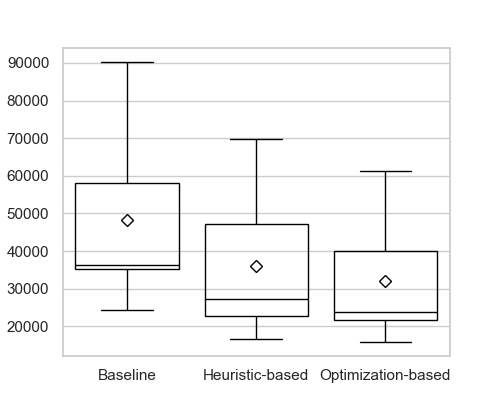}
    \caption{Costs.}
    \label{fig:Costs}
    \end{subfigure}
     \caption{Effectiveness of our baseline, heuristic-based and optimization-based solutions.}
    \label{fig:Effectiveness}
\end{figure*}
The costs are significantly reduced by our heuristic-based (-\savingHeuristicAbsolute$\approx$-\savingHeuristicPercentage) and optimization-based solutions (-\savingSMTAbsolute$\approx$-\savingSMTPercentage) compared to the baseline.
There are no cases in which the baseline outperforms our heuristic-based and optimization-based solutions.
The Wilcoxon signed-rank test~\cite{c4091bd3-d888-3152-8886-c284bf66a93a} confirms that our empirical evidence~($p<0.05$) about the superiority of the heuristic-based and optimization-based solutions compared to the baseline configuration.

The costs are significantly reduced (-\savingDiffHeuZAbsolute$\approx$-\savingDiffHeuZPercentage) by our optimization-based solution compared to the heuristic-based one.
There are no cases in which the heuristic-based outperforms the optimization-based solution. 
The Wilcoxon signed-rank test~\cite{c4091bd3-d888-3152-8886-c284bf66a93a} confirms our empirical evidence~($p<0.05$) about the superiority of the optimization-based solution compared to the heuristic-based solution.

\begin{Answer}[RQ1 --- Answer]
The heuristic-based and optimization-based solutions significantly reduced the average number of violated configuration rules (respectively -\diffConstrBrokenHeuBaseAbsolute$\approx$-\diffConstrBrokenHeuBasePercentage and -\diffConstrBrokenZBaseAbsolute$\approx$-\diffConstrBrokenZBasePercentage) and costs (respectively -\savingHeuristicAbsolute$\approx$-\savingHeuristicPercentage and -\savingSMTAbsolute$\approx$-\savingSMTPercentage) compared to the baseline.

The optimization-based solution significantly reduced the average number of violated configuration rules (-\diffConstrBrokenHeuZAbsolute$\approx$-\diffConstrBrokenHeuZPercentage) and costs (-\savingDiffHeuZAbsolute$\approx$-\savingDiffHeuZPercentage) compared to the heuristic-based approach.
\end{Answer}

\subsection{Efficiency (RQ2)}
\label{sec:Efficiency}

We assessed the efficiency of our solution as follows.

\begin{figure}
    \centering
    \includegraphics[width=\linewidth]{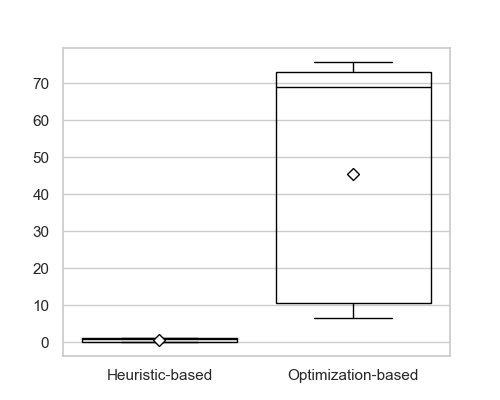}
    \caption{Seconds to compute a solution.}
    \label{fig:Efficiency}
\end{figure}

\textbf{Methodology}. We considered the experiments from RQ1. 
We compared the time the heuristic-based and optimization-based solutions required to find the configurations. Our experiments were executed on a Mac Studio equipped with an Apple M4 Max CPU and 36GB of unified memory running MacOS 26.0.1.

\textbf{Results}.  \Cref{fig:Efficiency} presents the time our heuristic-based and optimization-based solutions required to find the values for the configuration parameters. No time is reported for our baseline since engineers provided the baseline configuration values.

The average time required by our heuristic-based and optimization-based solutions is respectively \avgTimeToSolHeu (\textit{min}=\minTimeToSolHeu, \textit{max}=\maxTimeToSolHeu, \textit{StdDev}=\stdTimeToSolHeu) and \avgTimeToSolZ (\textit{min}=\minTimeToSolZ, \textit{max}=\maxTimeToSolZ, \textit{StdDev}=\stdTimeToSolZ) per extraction.
The time to find a solution is significantly lower in our heuristic-based solution compared to the optimization-based one, with an average difference of \diffTimeToSolHeuZAbsolute between the two.
There are no cases in which the optimization-based solution outperforms our heuristic-based solution. 
The Wilcoxon signed-rank test~\cite{c4091bd3-d888-3152-8886-c284bf66a93a} confirms that our empirical evidence~($p<0.05$) about the superiority of the heuristic-based solution compared to the optimization-based solution.

\begin{Answer}[RQ2 --- Answer]
Our results show that both the heuristic-based and the optimization-based solutions could find configuration values in practical time, respectively \avgTimeToSolHeu and \avgTimeToSolZ, on average, per extraction. The heuristic solution is considerably faster.
\end{Answer}

\subsection{Threats to Validity}
\label{sec:Threats}
Our results are subject to threats to internal and external validity.

\emph{Internal validity}.
The values we assigned to the constant values (those written in uppercase) of the rules from the catalogue (\Cref{sec:catalog}) threaten the internal validity of our results since other values may produce different results. 
To mitigate this threat, we have selected the constant values by consulting a domain expert.
The procedure used to compute the cost of a Kubernetes cluster threatens the internal validity of our results.
In practice, companies pay their resources depending on the number of nodes and not based on the workloads.
We consider the workload level since it provides a more precise estimation of our costs; Companies will likely overestimate the number of clusters and nodes they need.
Therefore, our results are penalized by our choice since considering clusters and nodes will increase the costs, and our approach would produce higher absolute savings. Relative savings, instead, are not impacted by our choice.

\emph{External Validity}.
The \numKubernetes Kubernetes we have selected and their initial configuration (the baseline) threaten the external validity of our results.
To mitigate this threat, we have selected a dataset that contains a considerable number of workloads (\numWorkloads) coming from clusters of different dimensions, scanned at various times of the day across different months. 
Moreover, these clusters come from real and important customers: They are not synthetic (as we explained in~\ref{sec:Benchmarks}). This further mitigates this threat to validity as it makes our dataset representative of what could be encountered in the field.

\subsection{Discussion and Lessons Learned}
\label{sec:Discussion}
We discuss the practical implications of our results, considering the violation of the configuration rules, costs, and time required to compute the configuration.
Then, we reflect on the comparison between the results of the heuristic-based and optimization-based solutions.

\emph{Violation of configuration rules}. Our solutions provide significant improvements in the configuration rules violated by the platform.
Specifically, the optimization-based solutions significantly reduced the average number of violated configuration rules (-\diffConstrBrokenZBaseAbsolute$\approx$-\diffConstrBrokenZBasePercentage).
This result has remarkable practical applications since the violation of configuration rules can have dramatic impacts on the safety of the application, especially when considering reliability-type rules. 
For example, \textsc{Insight} (the commercial software implementing parts of our work) would have prevented an outage that occurred in the infrastructure of a global leader in digital services and real-time transactional platforms in the retail market~\cite{akamasInsightPrevention}.
This outage was triggered by a dangerously high memory usage in a cluster of nodes. 
\textsc{Insight} would have recommended three actions: resizing memory requests, reducing overprovisioned CPU requests, and tuning the JVM heap sizes to prevent Out-Of-Memory incidents like the one that crippled the customer's infrastructure.

\emph{Costs}. Our solutions significantly reduce the costs.
Specifically, the optimization-based solution significantly reduced the costs (-\savingSMTAbsolute$\approx$-\savingSMTPercentage) compared to the baseline.
This result leads to an estimated annual average saving of \textsc{190\,488 USD}/mo per company, which is considerable.
This saving comes without any drawbacks for the safety of the application, since the configuration also violates fewer configuration rules.

\emph{Time Required to Compute the Configuration}. 
Both the heuristic-based and the optimization-based solutions required less than two minutes to find the configuration values (i.e., the maximum time required was 75.74 seconds).
This time is acceptable for early configuration phases of industrial applications.

\emph{Heuristic-based vs Optimization-based Solutions}.
Since the heuristic-based solution is hard-coded, we expected the configurations identified by it to satisfy more configuration rules compared to the configurations computed by the optimization-based solution.
Contrary to our expectations,  the optimization-based solution reduces both the costs and the number of violated configuration rules compared to the heuristic solver. 
Instead, the heuristic-based solution requires less computational time than the optimization-based solution, as expected.
However, the difference in the computational time is not a deal breaker since in both cases it is lower than two minutes (and therefore acceptable for customers). 
Furthermore, the optimization-based solver enables performance engineers to activate and deactivate configuration rules from the dashboard without the need to rewrite the source code. 
This provides greater flexibility in tailoring a solution according to the customer's demands.

Considering the remarkable results, \AKAMAS launched their latest product called ``\textsc{Insights}''~\cite{Insights} on November 10th, 2025 which implements  parts of our work.
\section{Related Works}
\label{sec:RelatedWork}
We discuss related works by considering approaches that  (a)~target general configuration problems, (b)~are tailored to Kubernetes, and (c)~are driven by requirements and constraints.

\emph{General Configuration Tuning}.
Configuration tuning is largely studied in the academic and industrial domains.
Manual configuration tuning is today ineffective~\cite{didona2015enhancing} due to the considerable number of parameters present in modern applications.
Genetic programming algorithms~\cite{10.1145/1712605.1712624, Gerasimou2018, 10500748, LEE2024103567},
autoscaling solutions based on control theory~\cite{baresi2020cocos} and queuing theory~\cite{lazowska1984quantitative}, have been widely employed and compared by recent works~\cite{10419899}.
Tuners leveraging a surrogate model (e.g., using Bayesian Optimization and its variants~\cite{8469102, jamshidi2016uncertaintyawareapproachoptimalconfiguration, cowenrivers2022hebopushinglimitssampleefficient, 10.1007/978-3-642-25566-3_40}) were also largely considered. 
Bayesian Optimization is also used by our industrial partner in one of its products, \AKAMAS~\cite{Akamas_product}, and is covered by the US Patent 11,755,451~\cite{stefano2023method} and Patent Application 17/707,498~\cite{cereda2022method}. 

Unlike these works, we propose domain-knowledge-driven configuration tuning and focus on Kubernetes. 

\emph{Kubernetes Configuration Tuning}.
Resource optimization and workload scheduling in container orchestration systems like Kubernetes is a significant area of research
(e.g.,~\cite{10.1145/3544788,app151810098,thota2023, 9139654, 11031045,10637606,amslaurea32040,8790136}). 
A recent review~\cite{11031045} analyzes the state of the art of optimization strategies for Kubernetes.
They spotlighted that current solutions often lack the integration of multi-dimensional metrics,
such as combining CPU utilization with memory usage,
network latency, and Quality of Service (QoS) metrics.
Our solution addresses these challenges by importing different configuration rules from the Catalog described in~\Cref{sec:catalog}.
The review classifies existing works into predictive scaling, reinforcement learning-based, deep learning-based, and multi-objective optimization-based approaches.
For example, existing approaches include adaptive scheduling algorithms for performance and cost optimization~\cite{app151810098}, AI solutions that rely on predictions, anomaly monitoring, queuing, and performance optimization to control resource scaling~\cite{thota2023}, an AI-based scaling engine that bases auto-scaling decisions on incoming requests~\cite{9139654}, and control-theoretical planners that manage container resources on-the-fly~\cite{baresi2021kosmos}.
Reinforcement learning has also been considered~\cite{10637606}. 

Our work on the optimization-based solver is related to these works and introduces a novel approach that leverages the Z3-SMT solver. 
Some works already exist where Z3 has been applied on Kubernetes: for instance~\cite{amslaurea32040} uses it to obtain a better autoscaler. 
A similar approach to the decision rules used by both our solutions is followed by ConfAdvisor~\cite{8790136}, which does not focus expressively on tuning the Kubernetes cluster, nodes, and workloads but rather provides advice on the whole IT configuration starting from a baseline. Moreover, ConfAdvisor uses its rules to find a good configuration, while our heuristic solution uses the rules only to identify problems, and our optimization-based solution tries to solve a linear system of equations based on these rules.

\emph{Configuration Tuning Guided by Requirements}.
Some tuners incorporate hard constraints expressing performance requirements to guide the configuration tuning using genetic algorithms~\cite{Gerasimou2018, LEE2024103567}. 
Some tools permit interactive configuration tuning, allowing the user to specify preferences between performance metrics~\cite{Cortellessa2025}.
Novel approaches based on the co-evolution of target an auxiliary requirements have also shown promising results~\cite{xiong2025cotunecoevolutionaryconfigurationtuning}.

Unlike these works, we provide performance engineers with the capability to select the configuration rules for the considered application.
After this configuration step is completed, our solution automatically finds an optimal configuration.
\section{Conclusion}
\label{sec:Conclusions}
This paper proposed a solution for finding configurations of industrial applications in a practical time, driven by domain knowledge.
Domain knowledge is provided via a portfolio of configuration rules developed by domain experts in similar applications over years of activity.
Performance engineers can design application-specific problems by importing from the catalog the configuration rules relevant to their problems.
We propose two alternative solutions to solve the optimization problems: heuristic-based and optimization-based.
We implemented these solutions within an industrial tool from a software company specializing in performance engineering. 

Our results confirm the effectiveness of our solution: compared to the baseline, the heuristic-based and optimization-based solutions provide significant average savings of \savingHeuristicAbsolute and \savingSMTAbsolute per cluster, respectively, accounting for \savingHeuristicPercentage and \savingSMTPercentage of the costs.
The solutions are found in practical time: the heuristic-based and optimization-based solutions required, respectively, \avgTimeToSolHeu and \avgTimeToSolZ seconds per extraction.
Our studies had a significant practical impact: Parts of our work have been integrated into the \textsc{Insights} commercial tool~\cite{Insights}, now commercialized by \AKAMAS. 

\section*{Data Availability}
The software developed within the context of this work can not be shared as it is the intellectual property of Akamas.
The authors can not share the evaluation data and the names of the Akamas customers these data refer to.

\section*{Acknowledgment}
\addcontentsline{toc}{section}{Acknowledgment}
The work of Claudio Menghi was in part supported by the European Union -- Next Generation EU. ``Sustainable Mobility Center (Centro Nazionale per la Mobilit\'a Sostenibile - CNMS)'', M4C2 -- Investment 1.4, Project Code CN\_00000023, and by projects SERICS (PE00000014) under the NRRP MUR program and GLACIATION (101070141).


\bibliography{bibliography}

\end{document}